\documentclass[a4paper,10pt]{article}
\usepackage[utf8x]{inputenc}
\usepackage{graphicx}
\usepackage{amsmath}
\usepackage[flushleft]{threeparttable}
\title{The new method to create Single Bubble Sonoluminescence}
\author{Jaroslav Anto\v{s}\footnote{retired from Institute of Experimental Physics, SAS, Kosice, Slovakia}}

\begin{document}

\maketitle

\begin{abstract}
 Sonoluminescence is a phenomenon which is known for some time, relatively easy to produce but still not fully
understood. A milestone was discovery of procedure for creating Single Bubble Sonoluminescene (SBSL) \cite{gaitan}
in 1989. Sonoluminescence in single bubble at well defined conditions, periodically produced for a long time, is
well suited for a systematic study. This procedure for SBSL was for many years considered as standard one for a study
this interesting phenomenon.\\
 About single bubble sonoluminescence was published wealth of sometime very surprising results, however
generally accepted solution to the problem is still missing.\\
 Another way (technique based on water hammer) to produce sonoluminescence was published in 2003 \cite{waterham}. By
 this method many orders of magnitude stronger signal  in comparison with standard procedure can 
be produced and also new phenomena have been observed.\\
 In this paper we describe another method to produce relatively stable SBSL. New technique is closely related to 
standard one. But it for example does not need degassing of working liquid and there is one more handle for a study
 of the sonoluminescence.
\newpage  
\end{abstract}
\newpage
\tableofcontents

\section{Introduction}
 Sonoluminescence - light produced under influence of sound is known for almost hundred years \cite{hist}.
 Random flashes of a light under influence of sound have been difficult for systematic study.
This kind of sonoluminescence is now called Multi Bubble Sonoluminescence (MBSL).\\
 About 30 years ago there was discovered a method\cite{gaitan} how to create stable sonoluminescence in a single bubble.
 This was really major step forward and there have been high expectations that Single Bubble Sonoluminescence (SBSL)
will be theoretically understood soon. But SBSL remains after more than 30 years from discovery \cite{gaitan} a puzzle. 
Topic is well covered in many original papers and reviews (see e.g. \cite{nature,brenn}). Dynamics of bubble
expanding and contracting in a field of ultrasonic standing wave is described reasonably well by
 Reyleigh - Plesset equation (see discussion e.g in \cite{brenn}). But
 there is still no generally
accepted theory to explain why short burst of light follows at right conditions contraction of bubble. There are many 
theories explaining some features but not full picture. Appart 
of fact of yet not fully theoretically understood burst of the light, spectrum of this light also represents a puzzle.
 It resembles spectrum of black body radiation at temperature $\approx$ 6000\textdegree - 20000\textdegree K (see e.g. \cite{spect} and
 references there). Continuous spectrum is an experimental fact and if interpretation of this spectrum as a blacbody radiation spectrum 
is correct,
and inside a bubble in very short time is such a  hot temperature, it rises a question - could this temperature be
at right condition  risen to even higher temperatures? Even to the temperatures at which nuclear fusion  can be observed?
 There is some theoretical support for positive answer to this question. And there is also positive answer to this
question by experiment \cite{taler}. However this claim has opponents in other qroups working on the topic which have not
been able to arrive to the same conclusion.\\
 Continuous spectrum represents a puzzle by itself. Why there are not seen spectral lines corresponding to elements inside gas or water?
 In multibubble sonoluminescence (MBSL) these spectral lines are observed.\\
Under special conditions (or environment) spectral lines are observed also in SBSL (\cite{specAr,specNaAr,specTb3}).\\
To get better insight into process of sonoluminescence one should be able to modify parameters
leading to sonoluminescence in wide range. It is established fact that sonoluminescence in water using 
standard procedure is achieved in range of amplitude of standing pressure wave in range of $\approx$ 120-140 kPa. Lower than 120 kPa amplitude
 is not sufficient to produce SL, in case of amplitude higher than 140 kPa stability of bubble is destroyed.  At single resonance frequency above 140 kPa bubble will disappear.
 There was found procedure to boost
slightly this parameter by adding higher harmonics to base resonance frequency \cite{boosted}.\\
Mechanism radically different from standard one to produce sonoluminescence was  presented around year 2003 \cite{waterham}. 
Sonoluminescence by this new mechanism was created in much larger bubbles, intensity of light was up to 4 orders
 of magnitude higher (than observed by standard method) and in some cases much larger time of pulse width was observed.\\
 Any new technique to produce
 sonoluminescence gives an opportunity to study this phenomenon from different point of view and to come closer
to the theoretical understanding of sonoluminescence.\\
In this paper we present yet another method to produce sonoluminescence and we hope that it will help 
to shed more light on this interesting phenomenon.\\
 Article is organized as follows. First used apparatus is briefly described. Then new procedure for creating a sonoluminescence
is explained. In measurement section observation of sonoluminescence by standard procedure are documented and
after that sonoluminescence by the new method is followed. In Summary results by both methods are compared and
 implication for future are contemplated.

\section{Apparatus}
\begin{figure}
 \includegraphics[width=8 cm]{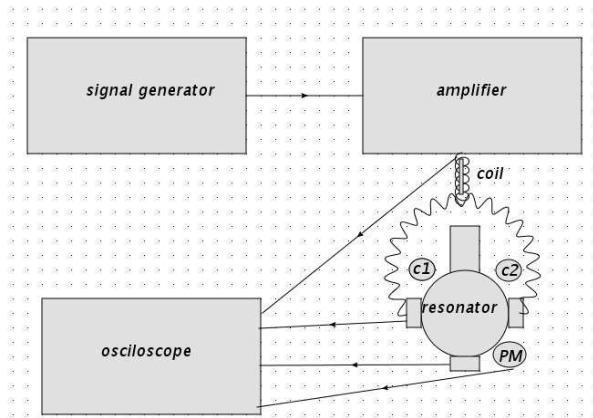}

 \caption{Schema  of our current apparatus  for a study of single bubble sonoluminescence. Resonator is equipped with piezoelements
on sides and microphone (small piezoelement)  bottom,
 C1,2 cameras, PM - photomultiplier }
  \label{schema}
\end{figure}

 Current apparatus consists of resonator, driving system, two cameras to view sonoluminescence from two different views and photomultiplier.
 Sensitivity of cameras used was claimed to be 0.0002 lux. \\
  Photomultiplier used in our measurements was M12FVC51 which is sensitive in spectral range 250-610 nm.\\
   Standard procedure to achieve single bubble sonoluminescence needs reducing gas dissolved in working liquid (degassing) in resonator. In this paper we use 
  as a working liquid distilled and deionized water. As a gas is used air.\\ 
  To control  level of degassing of liquid (in our case water) in resonator  we checked level of degassing  by a measurement
  dissolved oxygen. For this purpose was used commercial dissolved oxygen meter based on polarographic probe. Resolution of this device is 
  claimed to be 0.1 mg/l. We assume that reduction of oxygen by a given degassing procedure corresponds to reduction of dissolved air.\\
 Schematic view of apparatus is in Fig. \ref{schema}. Some results have been published in the past (see \cite{calibSL},\cite{spect}) using some elements of this kind
of apparatus.  
 \subsection{Resonator}
\begin{figure}
 \includegraphics[width=8 cm]{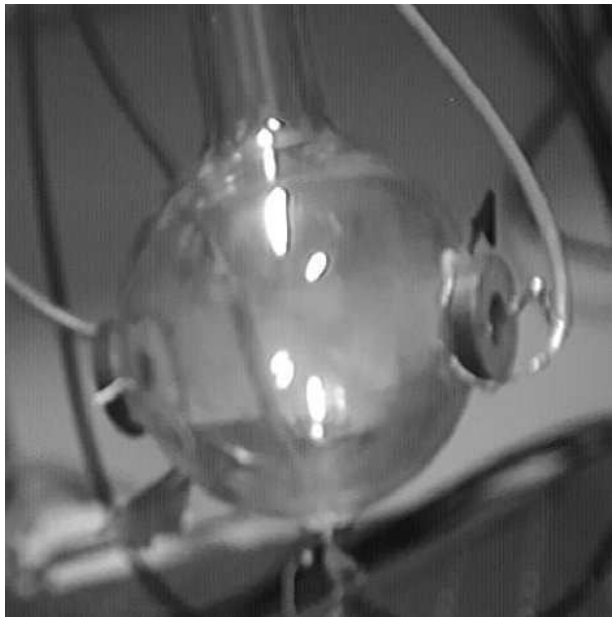}

 \caption{Picture of traditional resonator for study of single bubble sonoluminescence. There are two piezoelement
drivers glued opposite to each other at equatorial side of laboratory 100 ml round bottom flask. Their purpose is to produce
standing ultrasonic wave which will capture bubble at center. At the bottom there is small piezoelement (glued) which serves as microphone. }
  \label{res}
\end{figure}

 Traditional resonator for study of sonoluminescence (SL) looks like in Fig. \ref{res}. On sides one can see 2 piezoelements. Their purpose is
 to create standing ultrasonic wave in resonator filled by some liquid with small bubble at center. Bubble expands and shrinks in field of
 ultrasonic standing wave and at right conditions (amplitude and frequency of wave) it starts to periodically  burst light - phenomenon known as 
 single bubble sonoluminescence. At bottom of resonator there is a small piezoelement whose purpose is to detect response of the system.
 It works as a microphone and among other things it helps a great deal to tune system to achieve sonoluminescence.\\

 \subsection{Signal generator and audio amplifier}
  Minimum requirement on signal generator is stability on level of 1 Hz at frequency 20-30 kHz (for a given resonator), and amplitude up to 10 $V_{pp}$. 
  Signal of signal generator is amplified by audio amplifier. This signal is connected to a coil which forms with piezoelements on resonator
  a resonant LC circuit. Resonance condition for a given frequency is achieved by tuning inductance by moving a ferrite core of a coil 
  to a right position.
\subsection{Initiation of bubble for a Sonoluminescence}
 There are many ways how can be created  bubble at center of resonator which will under proper conditions lead to a
  SL. We tested
 several of them but finally we most of time use simple procedure by scratching surface of water in resonator. Bubbles 
 created this way at surface are attracted by ultrasonic field from surface to center of resonator. Provided
 right conditions are set. Scratching of water surface is realized by wire connected  to
 small electric motor. By a short electric pulse  wire will scratch surface of water,
 produces many bubbles and finally one bubble will stay at center of resonator and undergoes process leading
 to SL.
\subsection{Degassing procedure and control of level of degassed liquid (water)}
 Standard procedure to create single bubble sonoluminescence stresses importance of degassed liquid.
  There are several ways how to degas water. Simplest one is to boil water for a 15-20 minutes and then leave it to slowly cool down in a closed
  vessel. We tested this procedure however, apart of taking more time to have degassed water, also level of degassing was not well reproduced.
  Our choice was procedure based on using vacuum pump.
  By using vacuum pump  vacuum on level close to -90 kPa in a  given vessel with distilled deionized  water is created
  and at
  the same time water is stirred by magnetic stirrer. After 15 minutes of this procedure  dissolved air (oxygen) leaves water and
   amount of air  still present in water is measured by dissolved oxygen meter. At these conditions measured value is below 2 mg/l.
   These conditions are considered optimal for achieving SL characterized by high intensity and stability. 
   If water is degassed much below 2 mg/l created bubble (injected or by any possible way inserted into degassed water)
  tends to dissolve too fast and it is a little bit hard to tune conditions to create SL.\\

 \section{New procedure to create Single Bubble Sonoluminescence}
\begin{figure}
 \includegraphics[width=12 cm]{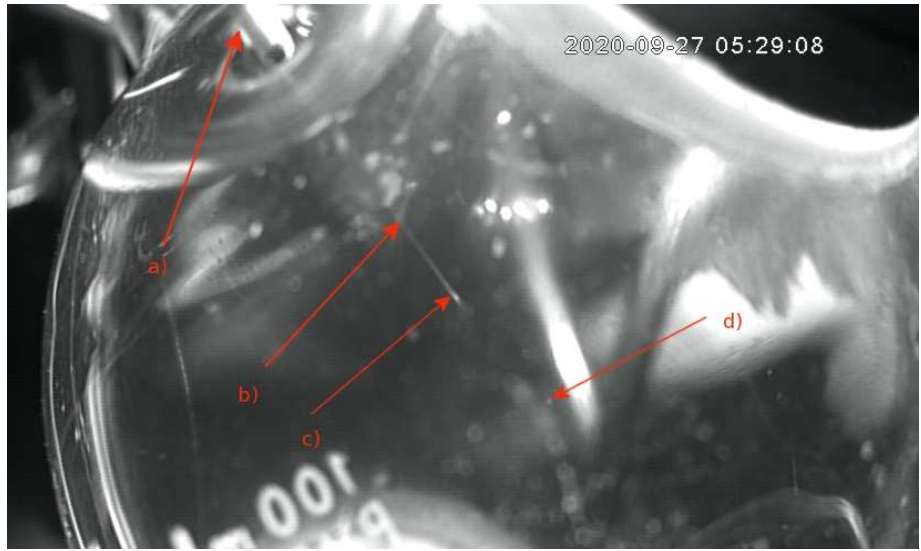}

 \caption{Ingredients of new method to produce SBSL. 
a)Points to  part of a narrow cylinder submerged in water. In ultrasonic field it initiate jet (presumably consisting of small
bubbles) to which points arrow marked as b).  c) points to head of jet. Arrow marked as d) points to small dot which is a bubble 
which at above conditions emit light - sonoluminescence. }
\label{nmech}
\end{figure}

  Standard method of creation of Single Bubble Sonoluminescence requires high level of reduction dissolved air in water
(or other liquid). Reason is that dissolved air in case of ultrasonic standing wave is attracted to the center of
 resonator and interfere with bubble already there. Process of expansion and compression leading to sonoluminescence
is disrupted. Problem with ``degassed'' liquid is to transfer bubble of gas to the center of resonator without
being absorbed by liquid. Trick is to set proper conditions (frequency and amplitude of standing wave) and create
multiple bubbles. One of them will be captured at center of resonator.
 Parameter space for sonoluminescence in this case (in water) is limited to a range about 120 kPa -140 kPa amplitude of pressure wave. There
 is a study \cite{boosted} which includes higher harmonics together with base frequency to create sonoluminescence and claims to move 
parameter space to higher values. \\ 
 For a new method no reduction of dissolved air in water is necessary. Just opposite a strong reduction of dissolved
gas will cause new method to fail to work.\\
 In Fig.\ref{nmech} 
there are shown main ingredients of the new method to produce SBSL. Picture was taken from a video which captured 
a dynamic process of creatin sonoluminescence by a new mechanism. Camera was rotated  45 $\deg$ in respect to resonator.\\
 Arrow a) in above figure points to a rod (diameter $\approx$ 2 mm) slightly immersed into water in resonator. This is most the
important ingredient.
 In a case of right frequency
and amplitude small bubbles are initiated around immersed rod  and they are accelerated to the center of resonator.  A jet is created
to which points arrow marked as b). Arrow c) points to head of jet where jet ends. Continuing supply of small bubbles from region where is narrow barrel
submerged ends here. It is observed some activity of bubbles in head of jet but there was observed no sonoluminescence.
 {\bf In some distance from the head of jet there is a small bubble which produces sonoluminescence}. There is observed some 
interactions between head of jet and  bubble. Most characteristic feature for this method is existence
of jet and correlated bubble which produces sonoluminescence.\\
 Better than description is to see a video clip \cite{newmech-videos}.
To achieve a given topology (jet and correlated bubble) is necessary tuning to find proper position of submerged rod,
proper frequency and amplitude of signal generator for driving piezoelements.

\section{Measurements}
 At first we repeat exercise with standard approach to single bubble sonoluminescence. We degas water by procedure
 described above.
 We start measurements with lowest level of dissolved air in water and gradually step by step we increase it. At each step we do
same set of measurements. Measurements consists of determination of dissolved oxygen at start and end of measurements,
establishing sonoluminescence by tuning frequency and amplitude of signal generator, recording signals from photomultiplier
by oscilloscope (together with voltage and current - input to LC circuit created by coil and driving piezoelements). \\
Titles of subsections are composed of parameters of signal generator at which sonoluminescence was achieved and
average of dissolved oxygen level at start and end of measurements. At low levels of DO there is significant
 increase of dissolved oxygen in water during measurement.\\
 \subsection{Standard method}
\subsubsection{SL f=27.35 kHz, A=9.6 $V_{pp}$ DO=2.9 mg/l}
 We started a measurement at level dissolved oxygen in water (DO) 2.3 mg/l and at the end of measurements
result was 3.5 mg/l. Stable sonoluminescence was found at frequency f=27.35 kHz and amplitude of signal generator
A=9.6-10 $V_{pp}$. As expected at this condition gas bubbles tend to be quickly dissolved therefore new method is
not expected to work and it was tested that it did not work.\\
\subsubsection{SL f=27.374 kHz, A=8.49 $V_{pp}$, DO=4.1 mg/l}
 DO was increased from 3.5 mg/l to 4.0 mg/l. Stable SL was found at parameters f=27.374 kHz, A=8.49 $V_{pp}$.
 After measurements DO was measured and observed value was 4.2 mg/l.
\subsubsection{SL f=27.374 kHz, A=8.49 $V_{pp}$, DO=4.85 mg/l }
 DO was increased from 4.2 mg/l to 4.7 mg/l. Stable SL was found at the same conditions as in previous case.
 After measurements DO was measured and observed value was 5 mg/l.
\subsubsection{SL f=27.374 kHz, A=7.52 $V_{pp}$, DO=5.7 mg/l}
 DO was increased from 5 mg/l to 5.7 mg/l. Stable SL was observed at the same frequency and slightly lower
amplitude of signal from signal generator.
After measurements measured DO was the same as starting one (5.7 mg/l).
\subsubsection{Summary of SL based on standard method}
\begin{figure}
%\begin{center}
$
 \begin{array}{cc}
  
  \includegraphics[width=6 cm]{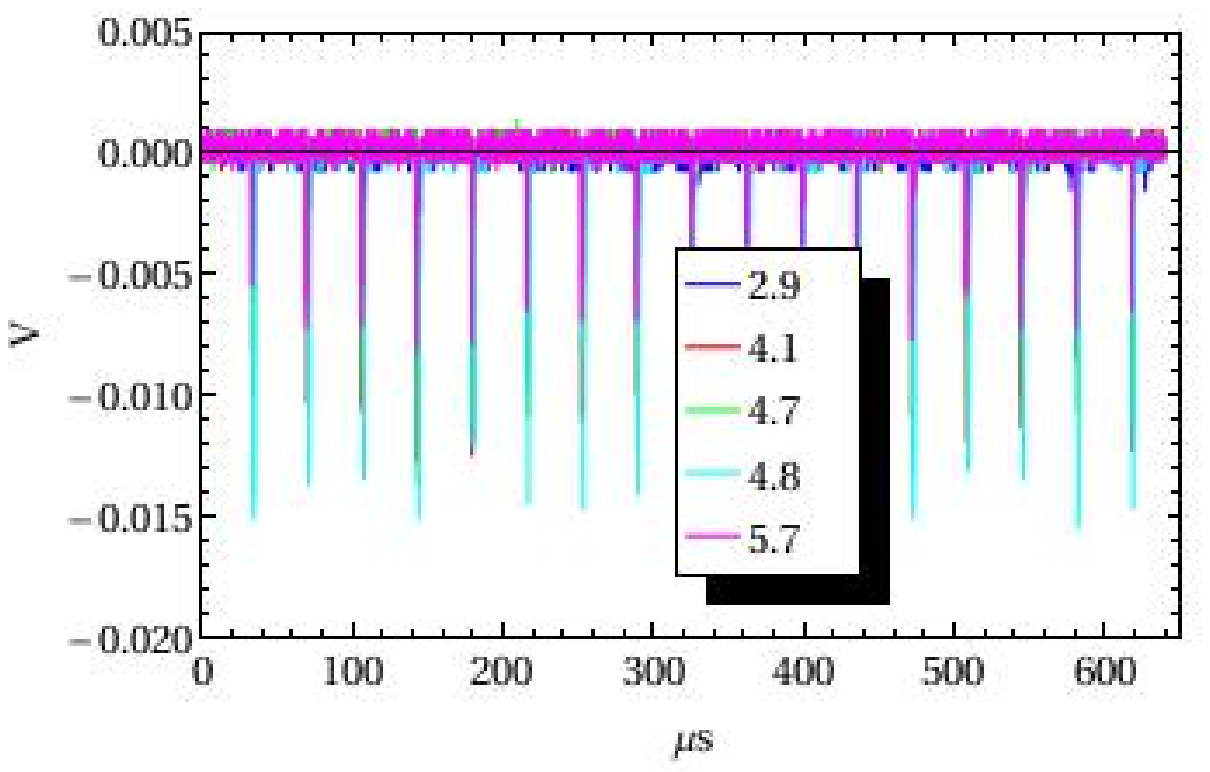} &
\includegraphics[width=6 cm]{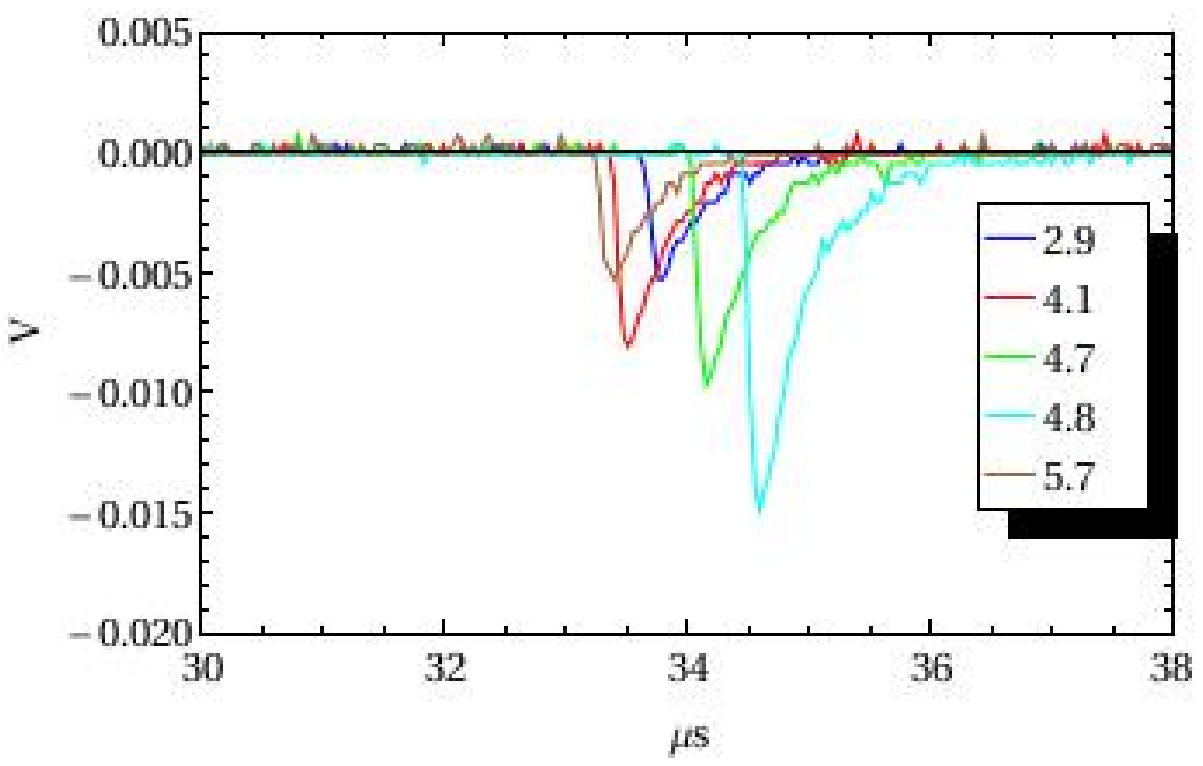}

\end{array}$
%\end{center}
 \caption{Photomultiplier response to a sonoluminescence created by standard method. Numbers at legend mean
average of dissolved oxygen level in water inside resonator for a given measurement. Plot on left side demonstrates uniform response of SL
over $\approx$ 600 $\mu s$. On  the right side details of individual response to SL is emphasized. }
  \label{oldmech}
\end{figure}

 Results are summarized in Fig. \ref{oldmech}. Sonoluminescence signal represents short burst of light per period. 
Uniform and periodically repeated. Width of signal is limited by resolution of a given photomultiplier.
A bubble producing SL is spatially fixed at the center of resonator.

\subsection{New method}
 At level of DO $\approx 6 mg/l$ was problem to establish sonoluminescence by standard method. SL by standard method
was still achieved but of very low intensity. Conditions (large level of dissolved air in water) 
 started to favor new method. 
\subsubsection{SL f=27.265 kHz, A= 10 $V_{pp}$, DO=6.5 mg/l}
 DO at start 6.2 mg/l.\\
 SL was achieved also using standard method at f=27.374 kHz, A=8.44 $V_{pp}$ but not stable and very low intensity.
 By new method stable SL was achieved at wide range of frequencies f=27.265 - 27.424 kHz (A=10 $V_{pp}$).
 Strongest signal was observed at f=27.265 kHz, A=10 $V_{pp}$.\\
 DO at the end of measurements 6.8 mg/l.
\subsubsection{SL f=27.385 kHz, A=7.07 $V_{pp}$, DO=7.4 mg/l}
\begin{figure}
$\begin{array}{cc}
 \includegraphics[width=6 cm]{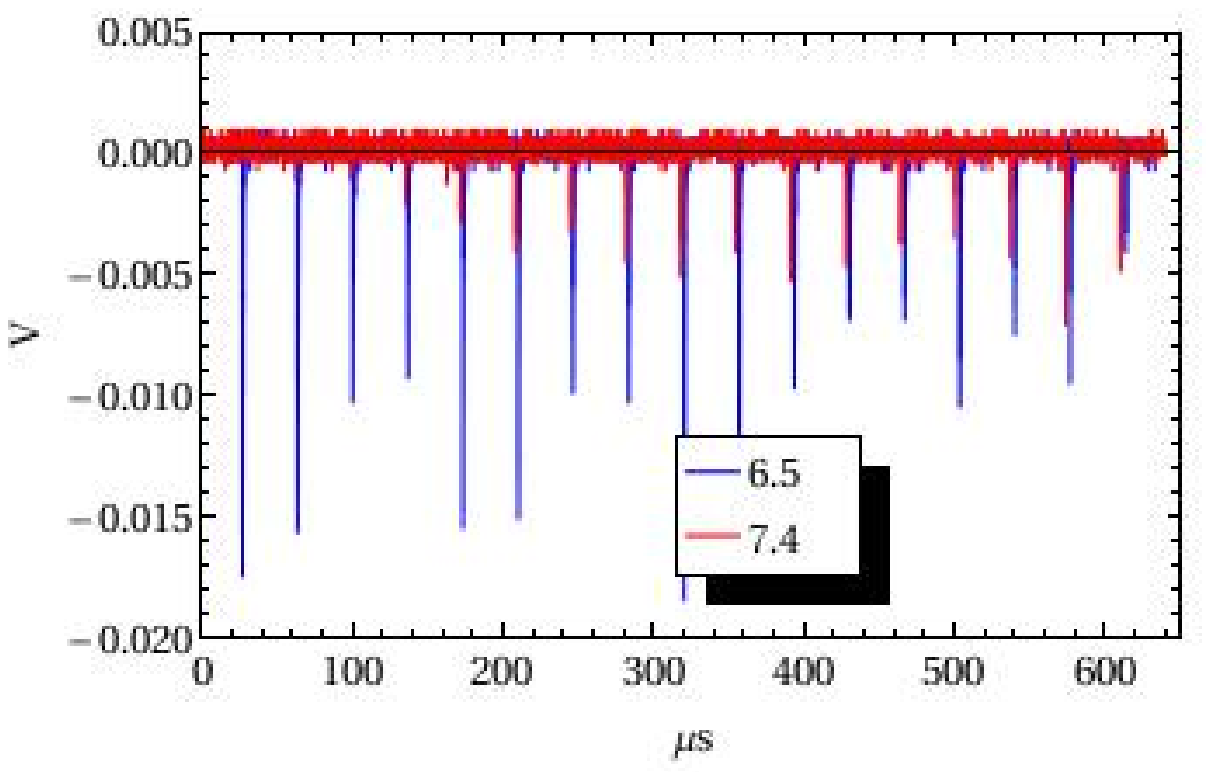} &
\includegraphics[width=6 cm]{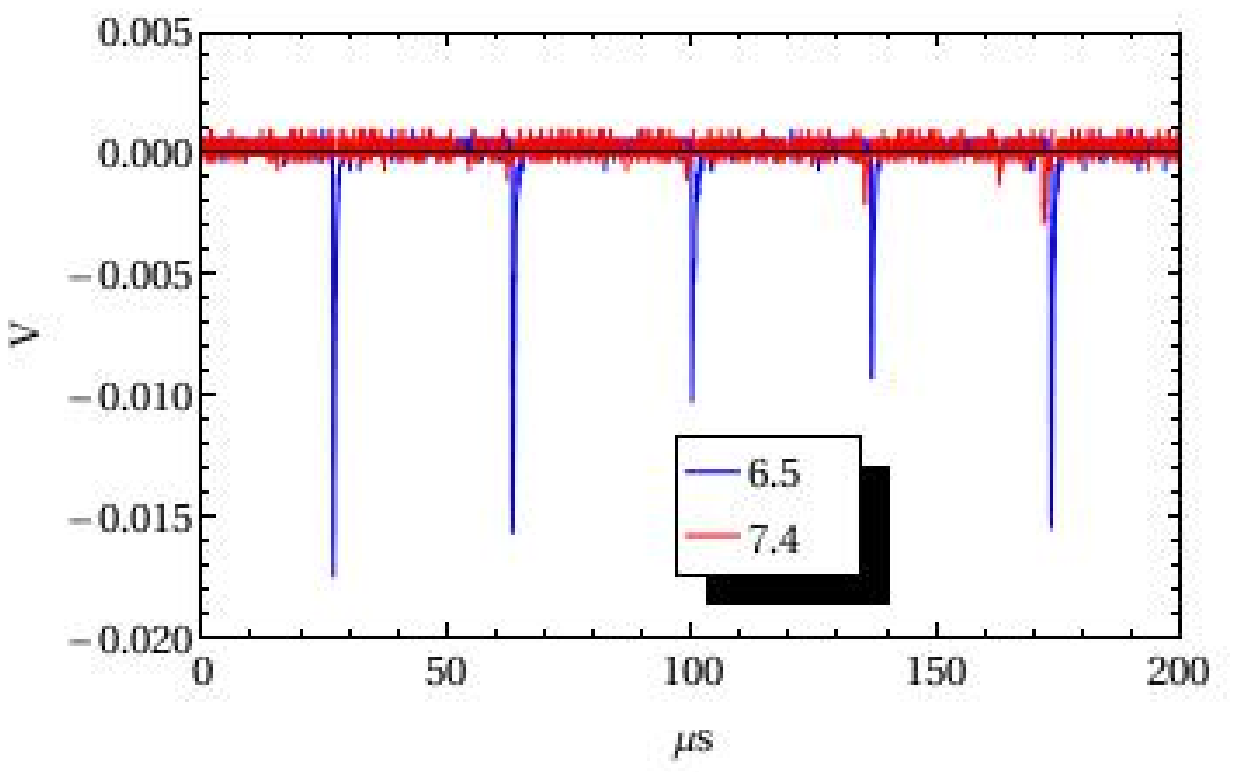} \\
 \includegraphics[width=6 cm]{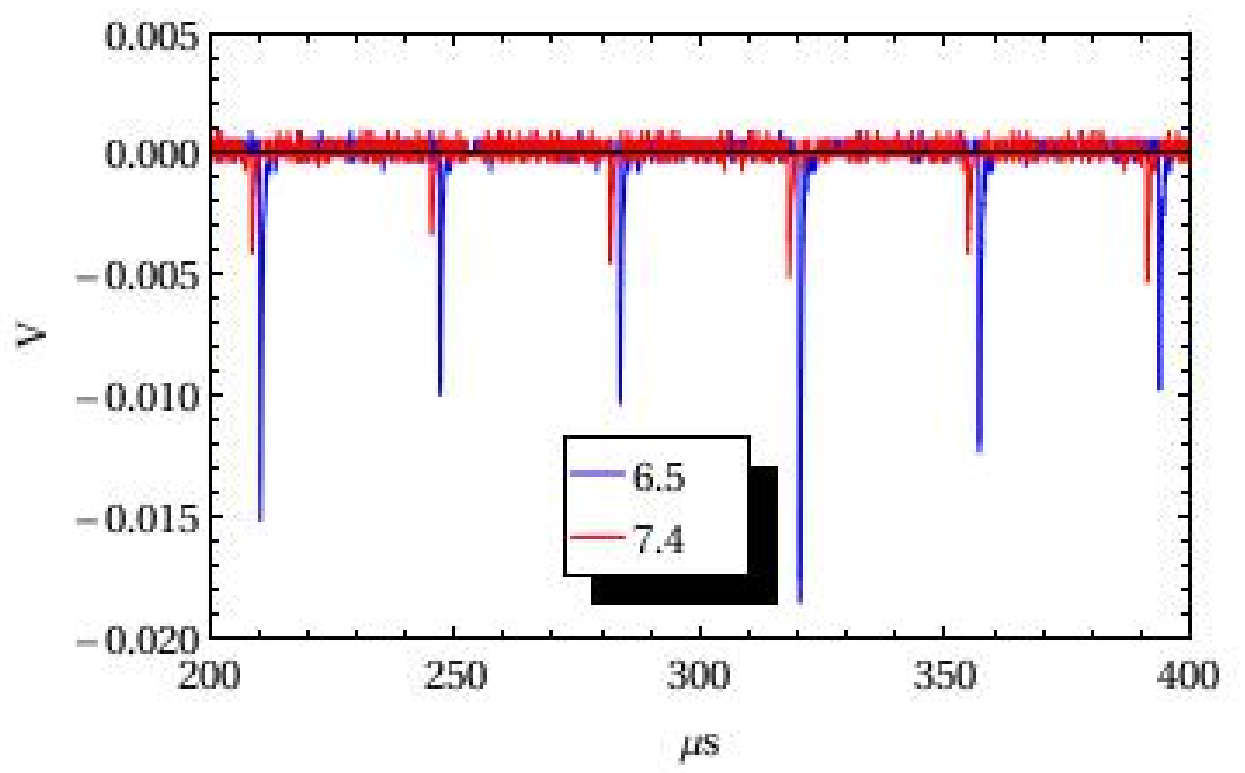} &
\includegraphics[width=6 cm]{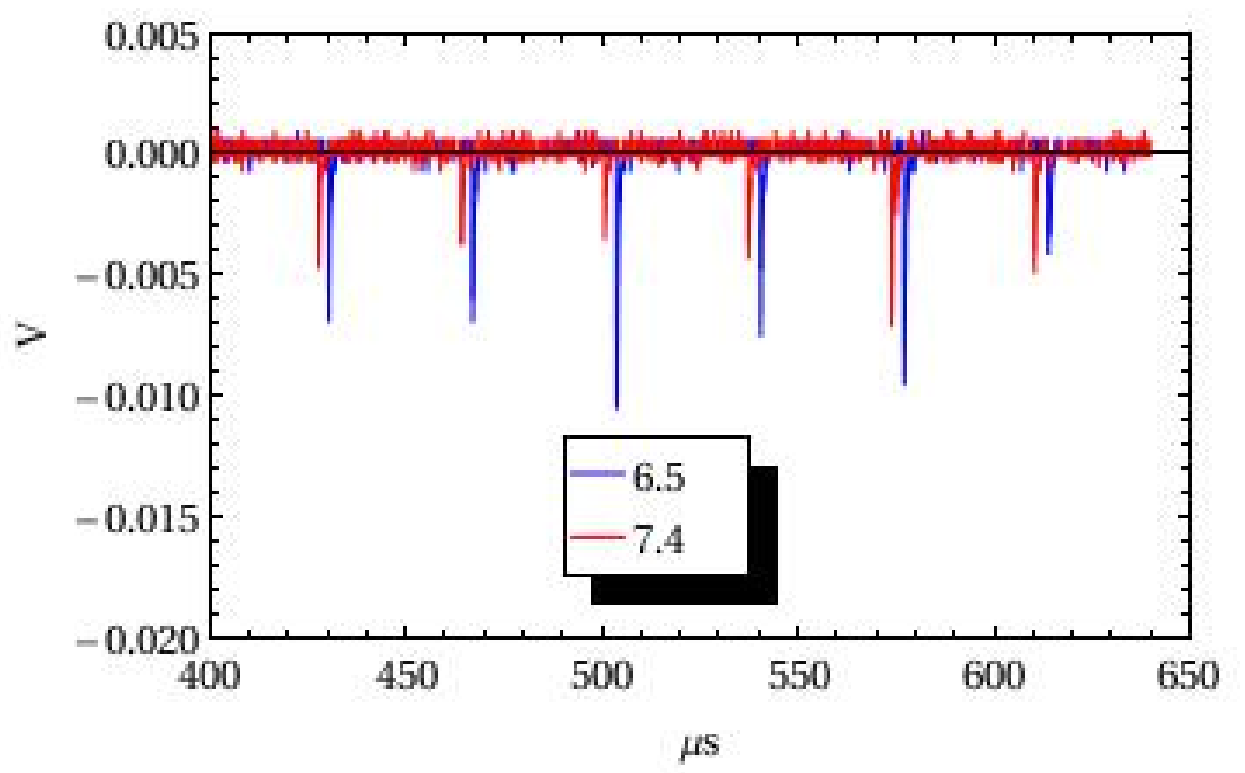} \\
\end{array}$
 \caption{New method for creating SL. Upper left plot demonstrates that SL by new mechanism has the same features as
SL created by standard method. Short burst of light periodically repeated. However as one can see 
 signal is not so uniform as in case of standard method  and it can be even missing for several periods.}
  \label{newmechDO}
\end{figure}

 DO was increased to 7.4 mg/l.SL was established at f=27.35 kHz A=10 $V_{pp}$ but stable and stronger
signal was observed at f=27.385 kHz and A=7.07 $V_{pp}$.
 Above two measurements are summarized in Fig. \ref{newmechDO}.
\subsubsection{DO= 7.7-8.1 mg/l}
\begin{figure}
$\begin{array}{cc}
\includegraphics[width=6 cm]{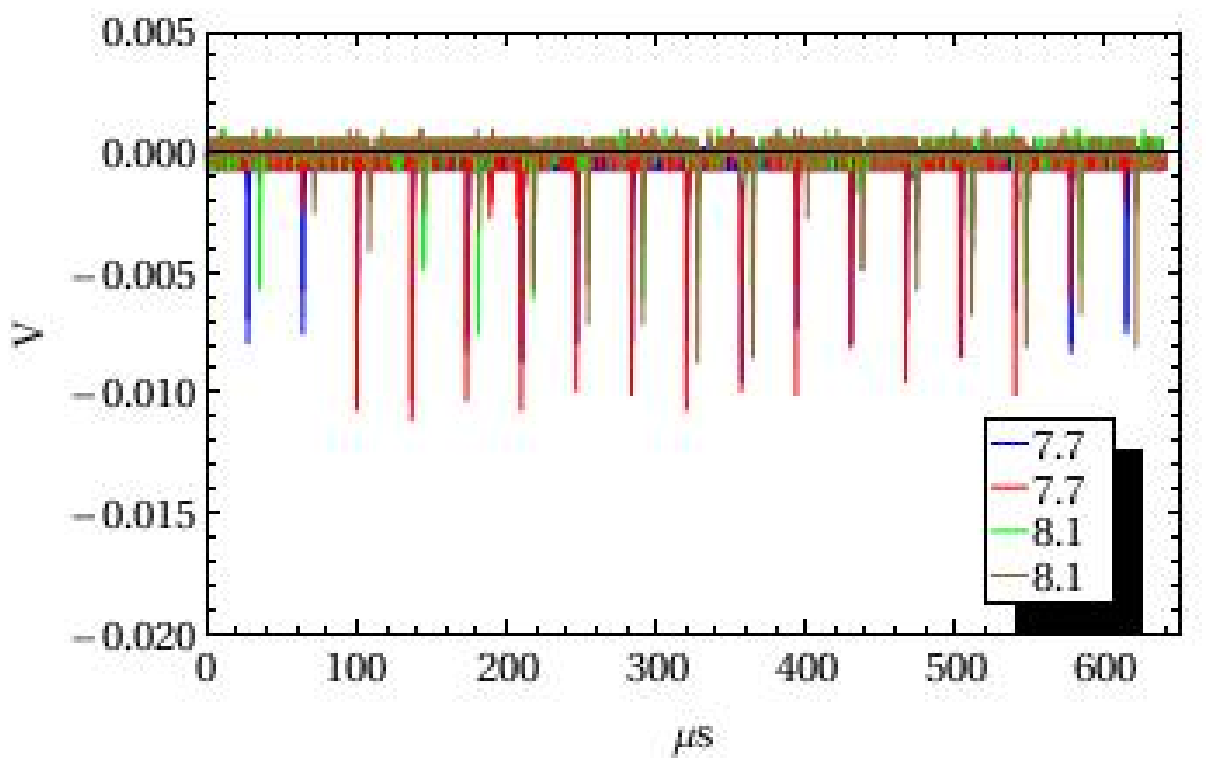} &
\includegraphics[width=6 cm]{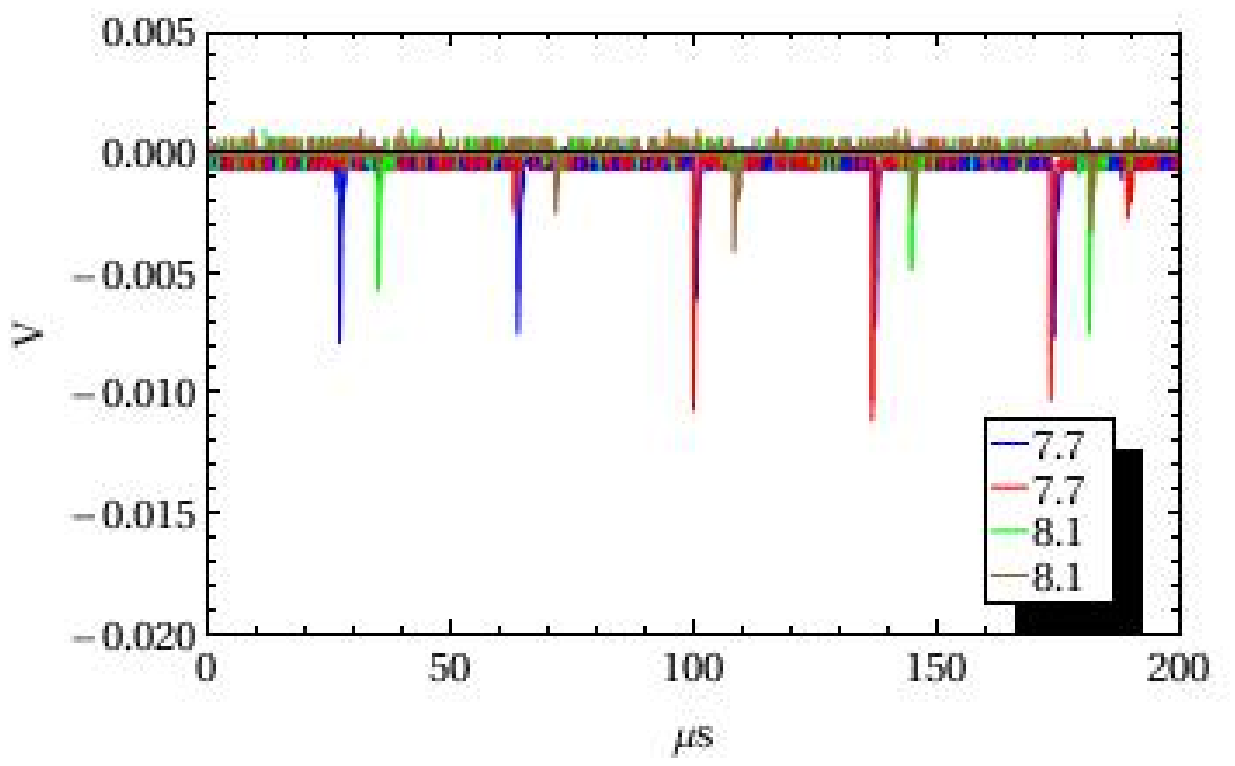} \\
 \includegraphics[width=6 cm]{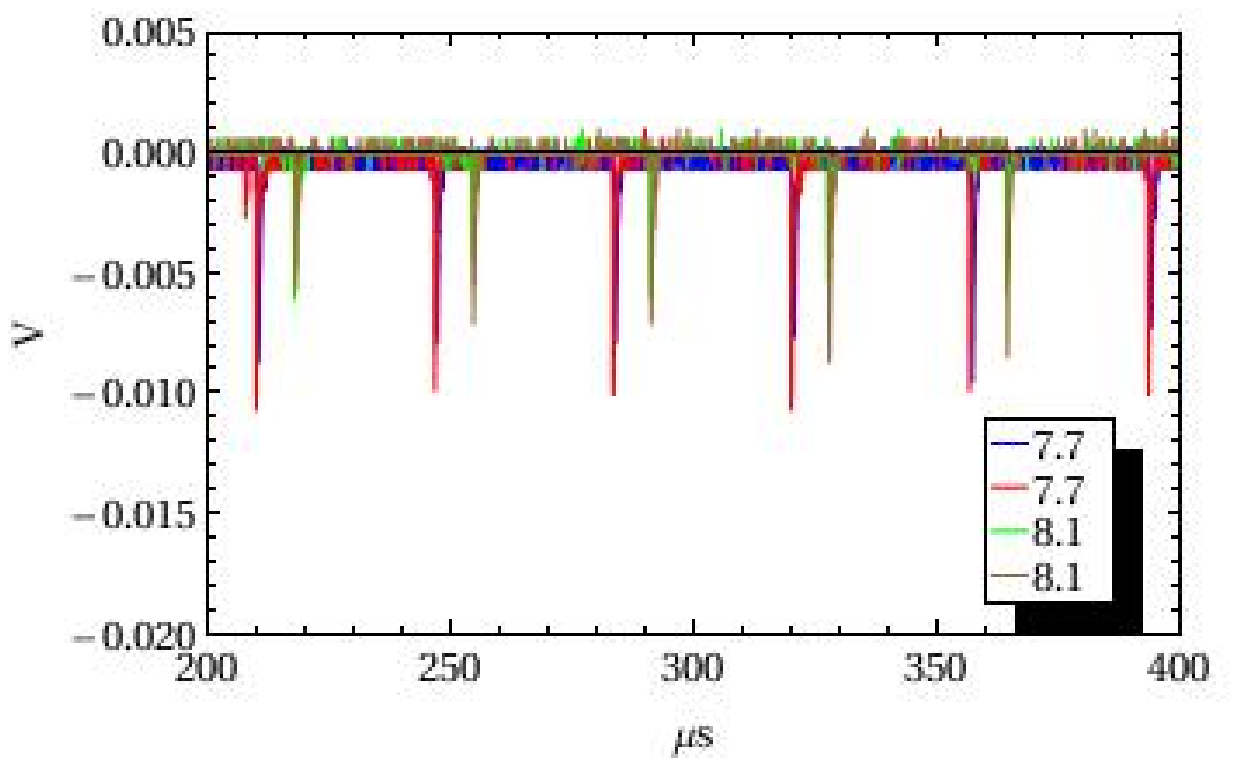} &
\includegraphics[width=6 cm]{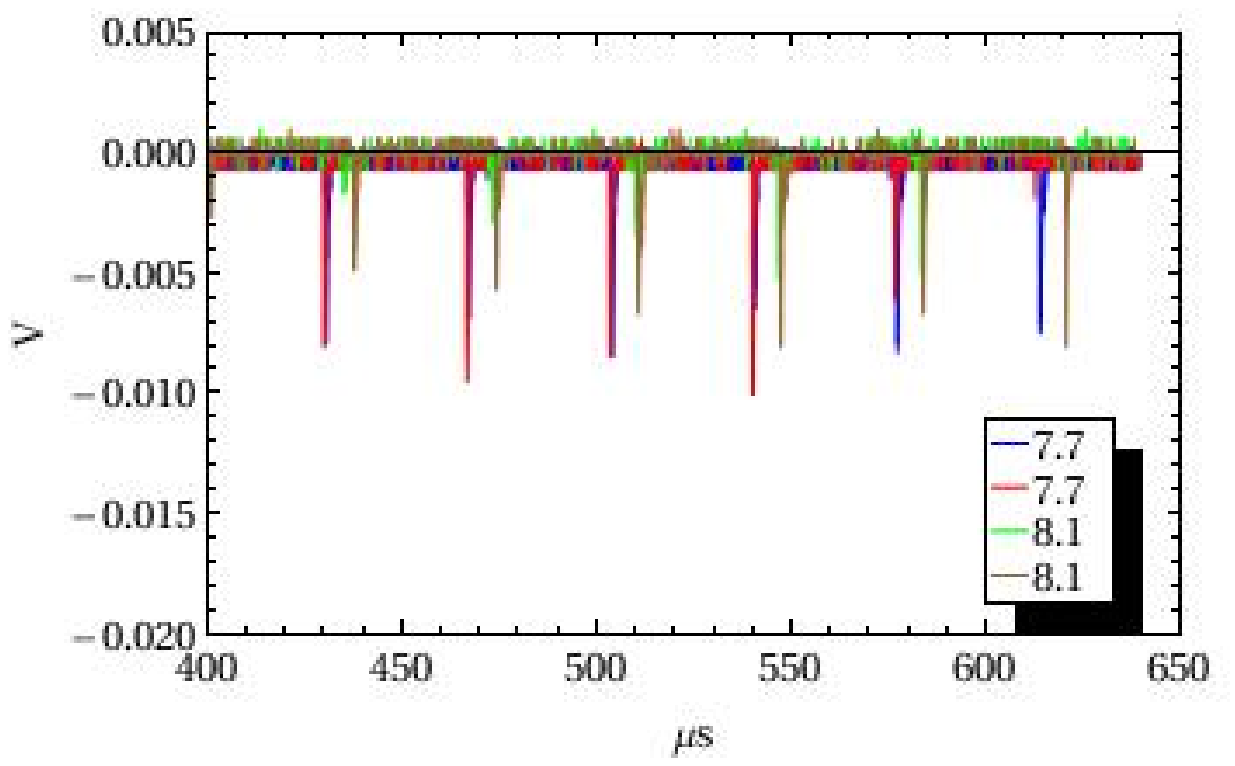} \\
\
\end{array}$
 \caption{New method for creating SL was used. Same features as in Fig. \ref{newmechDO} are observed. }
  \label{newmechDO2}
\end{figure}

 DO was increased to 7.7 mg/l and SL was established by new mechanism at
conditions:\\
\begin{itemize}
 \item f=27.278 kHz, A=6.54 $V_{pp}$, DO=7.7 mg/l
\item f=27.270 kHz, A=6.34 $V_{pp}$, DO=7.7 mg/l
\item f=27.344 kHz, A=10 $V_{pp}$, DO=8.1 mg/l
\end{itemize}
 Results are summarized in Fig. \ref{newmechDO2}
 
\clearpage
\section{Summary}

\begin{figure}
$\begin{array}{cc}
 \includegraphics[width=6 cm]{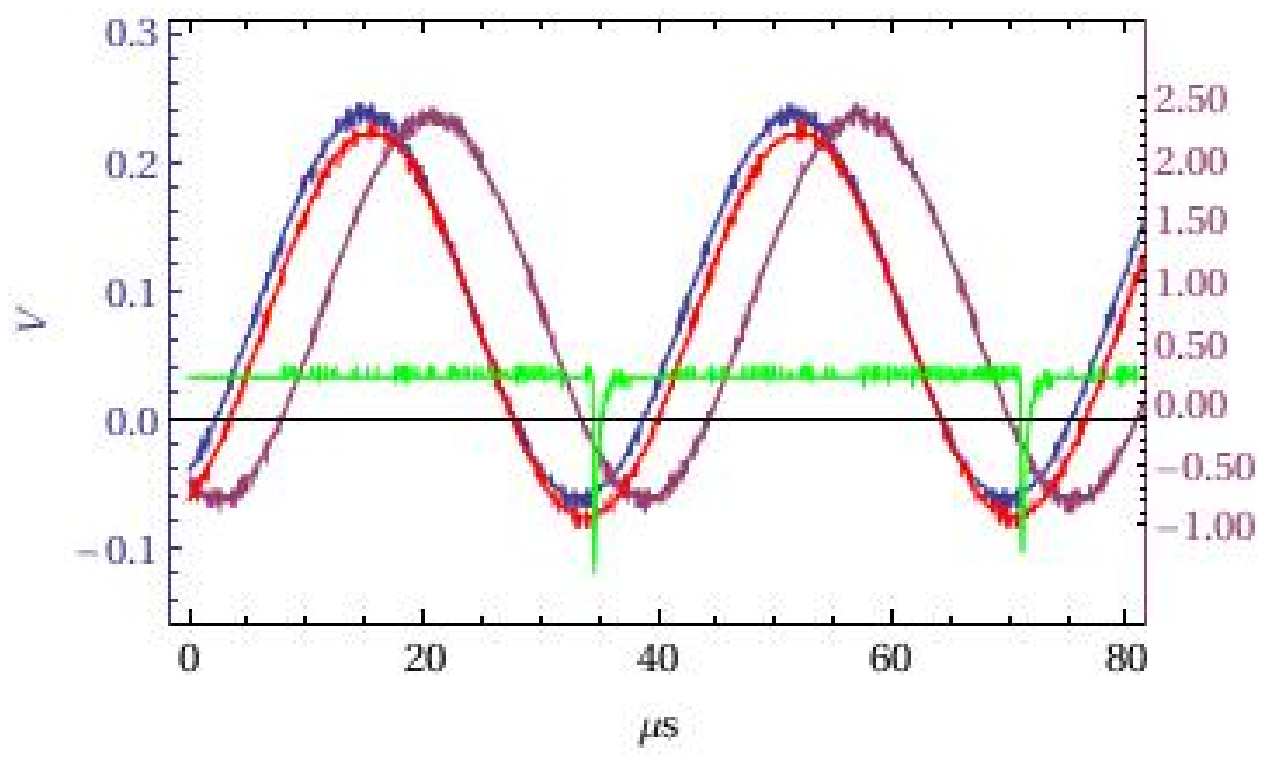} &
\includegraphics[width=6 cm]{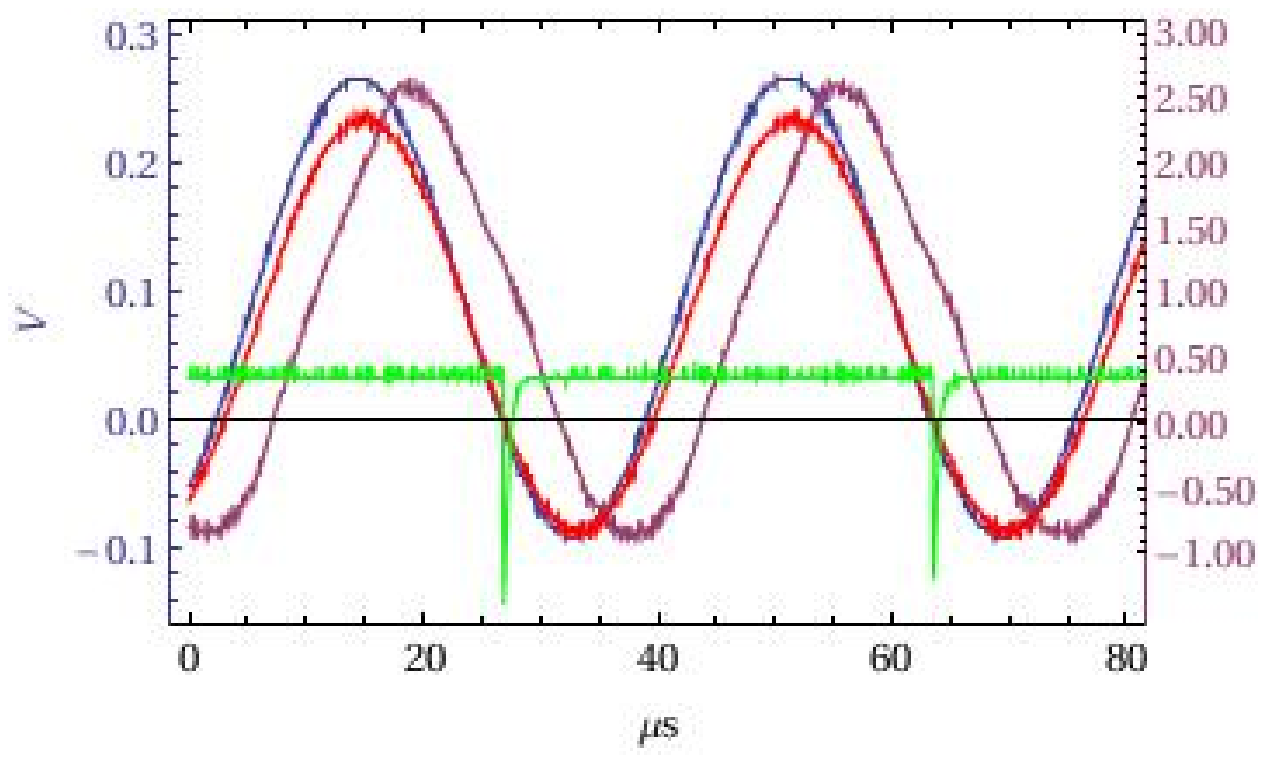} \\
 
\end{array}$
 \caption{A comparison of sonoluminescence by standard mechanism (left) and new one (right). In plots are included oscilloscope traces
corresponding to voltage applied to LC circuit (blue), current (red) and microphone signal (brown).
 Photomultiplier trace (green) is scaled
by factor 10 to be well described in 2 units scales used in plot. Left side scale is for blue,red
 and green trace.}
  \label{nmech-sl}
\end{figure}

 In Fig. \ref{nmech-sl} are displayed several standard plots with photomultiplier trace included. Plot on the left
 demonstrates SL in case of standard method on the right SL created by new method. As an example are choosen
strongest signals in both cases. From this comparison one tends to conclude that SL in both cases is close (the same). 
However for final conclusion one needs to measure and compare other parameters (like spectrum, pulse width).
 By a comparison of standard and new mechanism of production single bubble sonoluminescence we arrive at 
following differences and similarities:\\
\begin{itemize}
 \item short burst of light and periodic repeating of signal in both cases. 
Signal is about the same order of magnitude. In case of standard method seems to be more stable.\\
\item spatial position of bubble is in case of standard method is stable. 
In case of new method bubble is moving.
\end{itemize}
 In this paper we demonstrate a method to produce SBSL based on jet and correlated bubble. One should mention that by 
inserting rod into resonator depending on parameters (like how deep, frequency, amplitude) there are visible
also other topologies which also produce somewhat random sonoluminescence. Some topologies have
 been observed to produce even stronger spikes of sonoluminescence but their appearance seemed to be unrelated to a basic
driving frequency. These topologies have not been systematically studied yet.\\
Jet and correlated bubble topology provides straightforward mechanism for a production of SBSL.

\end{document}